\documentclass{elsart}

\usepackage{graphicx}
\usepackage{amssymb}

\newcommand{\CP}{\mbox{\it CP}}
\newcommand{\ra}{\mbox{~$\rightarrow$}~}

\newcommand{\gp}{\mbox{$p$}}

\newcommand{\gn}{\mbox{\it n}}

\newcommand{\gL}{\mbox{$\Lambda$}}

\newcommand{\Xim}{\mbox{$\Xi^-$}}

\newcommand{\Xiz}{\mbox{$\Xi^0$}}

\newcommand{\Omm}{\mbox{$\Omega^-$}}

\newcommand{\Sigp}{\mbox{$\Sigma^+$}}
\newcommand{\Sigm}{\mbox{$\Sigma^-$}}

\newcommand{\gam}{\mbox{$\gamma$}}

\newcommand{\piz}{\mbox{$\pi^0$}}

\newcommand{\pip}{\mbox{$\pi^+$}}
\newcommand{\pim}{\mbox{$\pi^-$}}

\newcommand{\Km}{\mbox{$K^{-}$}}

\newcommand{\alL}{\makebox{$\alpha_{\Lambda}$}}

\newcommand{\alXi}{\makebox{$\alpha_{\Xi}$}}

\newcommand{\alOm}{\makebox{$\alpha_{\Omega}$}}

\begin{document}

\begin{frontmatter}

\title{\centerline{Observation of Parity Violation in the
                   {\boldmath$\Omm\ra\gL\Km$} Decay} 
}

\author[uva]{L.C. Lu},
\author[iit]{R.A. Burnstein},
\author[iit]{A. Chakravorty},
\author[as]{Y.C. Chen},
\author[ucb]{W.-S. Choong},
\author[usa]{K. Clark},
\author[uva]{E.C. Dukes\corauthref{cor}},
\author[uva]{C. Durandet},
\author[ug]{J. Felix},
\author[lbl]{Y. Fu},
\author[lbl]{G. Gidal},
\author[umich]{H.R. Gustafson},
\author[uva]{T. Holmstrom},
\author[uva]{M. Huang},
\author[fnal]{C. James},
\author[usa]{C.M. Jenkins},
\author[lbl]{T.D. Jones},
\author[iit]{D.M. Kaplan},
\author[umich]{M.J. Longo},
\author[iit]{W. Luebke},
\author[ucb,lbl]{K.-B. Luk},
\author[uva]{K.S. Nelson},
\author[umich]{H.K. Park},
\author[ul]{J.-P. Perroud},
\author[iit]{D. Rajaram},
\author[iit]{H.A. Rubin},
\author[fnal]{J. Volk},
\author[iit]{C.G. White},
\author[iit]{S.L. White},
\author[lbl]{P. Zyla}

\address[as]{Academia Sinica, Nankang, Taipei 11529, Taiwan, ROC}
\address[ucb]{University of California at Berkeley, Berkeley, CA 94720,
              USA}
\address[fnal]{Fermi National Accelerator Laboratory, Batavia, IL 60510,
               USA}
\address[ug]{Universidad de Guanajuato, 37000 Le\'{o}n, Mexico}
\address[iit]{Illinois Institute of Technology, Chicago, IL 60616, USA}
\address[ul]{Universit\'{e} de Lausanne, IPHE, CH-1015 Lausanne, Switzerland}
\address[lbl]{Lawrence Berkeley National Laboratory, Berkeley, CA 94720,
              USA}
\address[umich]{University of Michigan, Ann Arbor, MI 48109, USA}
\address[usa]{University of South Alabama, Mobile, AL 36688, USA}
\address[uva]{University of Virginia, Charlottesville, VA 22901, USA}

\corauth[cor]{Corresponding author.
Tel.: 1-434-982-5364, fax: 1-434-982-5375;
{\it E-mail address:} \texttt{craigdukes@virginia.edu} (E.C. Dukes).}
\mbox{}\\[0.1in]
\centerline{HyperCP Collaboration}

\begin{abstract}
The $\alpha$ decay parameter in the process \Omm\ra\gL\Km\ has been 
measured from a sample of 4.50 million unpolarized \Omm\ decays
recorded by the HyperCP (E871) experiment at Fermilab and found to be 
$[1.78{\pm}0.19({\rm stat}){\pm}0.16({\rm syst})]{\times}10^{-2}$.
This is the first unambiguous evidence for a nonzero $\alpha$ decay 
parameter, and hence parity violation, in the \Omm\ra\gL\Km\ decay. \\[0.2in]
\noindent
{\it PACS:}
11.30.Er, 
13.30.Eg, 
14.20.Jn  \\
\noindent
{\it Keywords:} Omega-minus decays; Parity violation; Alpha decay parameter
\end{abstract}

\end{frontmatter}

%\section{Introduction}
Our knowledge of the \Omm\ hyperon and its decays remains incomplete,
despite its long and illustrious role in particle physics.
Its spin and parity have not been firmly established,\footnote{The 
\Omm\ spin has not yet been determined, but measurements
have ruled out $J = \frac{1}{2}$ and are consistent with
the quark-model prediction of $J = \frac{3}{2}$; see \cite{omega_spin}.
Throughout this Letter we assume that the \Omm\ is spin-$\frac{3}{2}$.
}
and it alone among the hyperons has yet to
exhibit parity violation in its two-body weak decays.
The Particle Data Group (PDG) values of the
$\alpha$ decay parameters of the three such decays,
\Omm\ra\gL\Km, \Omm\ra\Xim\piz, and \Omm\ra\Xiz\pim,
respectively $-0.026{\pm}0.023$, $+0.05{\pm}0.21$,
and $+0.09{\pm}0.14$ \cite{pdg}, are consistent with zero,
where $\alpha$ is a measure of the interference between the $P$- and
$D$-wave final-state amplitudes:
\begin{equation}
   \alpha = \frac{2{\rm Re}(P^{\ast}D)}{|P|^2 + |D|^2}.
\end{equation}
A nonzero value of $\alpha$ is manifest evidence of parity violation.
In contrast, all other hyperons have been shown to
have nonzero $\alpha$ decay parameters.
The smallest are those of the \Sigp\ra\gn\pip\ and the
\Sigm\ra\gn\pim\ decays, both of which are 0.068;
the largest is almost unity: $\alpha = -0.980$ in \Sigp\ra\gp\piz\ decays
\cite{pdg}.
The two-body nonleptonic \Omm\ decays are expected to be nearly parity
conserving \cite{Finjord}, and hence predominantly $P$ wave,
implying a small $\alpha$ decay parameter, which is consistent with
the experimental results.

Recently, we have reported evidence of parity 
violation in an analysis of 0.96 million \Omm\ra\gL\Km\ decays taken
in the 1997 Fermilab fixed-target running period, yielding
$\alOm = [2.07{\pm}0.51({\rm stat}){\pm}0.81({\rm syst})]{\times}10^{-2}$
\cite{hypercp97}. (Throughout this Letter \alOm\ will 
refer only to the \gL\Km\ decay mode of the \Omm.)  
We report here another measurement of \alOm\
using 4.50 million events taken during the
1999 Fermilab fixed-target running period.

\begin{figure}[htbp]
\centerline{\includegraphics[width=5.0in]{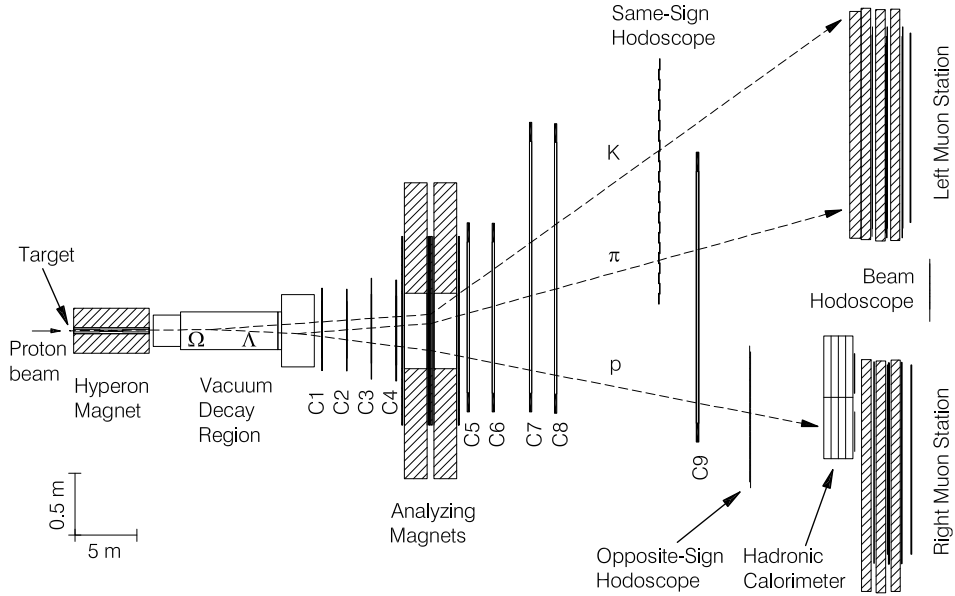}}
\caption{Plan view of the HyperCP spectrometer.
  \label{fig:spect}}
\end{figure}

%\section{The HyperCP Detector}
The experiment was mounted in the Meson Center beam line
at Fermilab using an apparatus \cite{apparatus} 
built to search for \CP\ violation in hyperon decays 
(see Fig.~\ref{fig:spect}). 
A negatively charged secondary beam with an
average momentum of 160\,GeV/$c$ was
produced by steering an 800\,GeV/$c$ proton beam onto a 
60\,mm long, $2{\times}2$\,mm$^2$ wide, Cu target.  The target was followed by
a curved collimator embedded in a dipole magnet (``Hyperon Magnet'').
The \Omm's were produced at an average angle
of $0^{\circ}$, ensuring that their polarization was zero.
The secondary beam exited the collimator upward at 
19.51\,mrad relative to the incident proton beam direction.
A 13\,m long evacuated pipe (``Vacuum Decay Region'') immediately 
followed the collimator exit.
After the Vacuum Decay Region was a high-rate
magnetic spectrometer employing nine multiwire proportional
chambers (MWPCs), four in front of a pair of dipole magnets (``Analyzing
Magnets''), and five behind.
Particles with the same sign as the secondary beam were deflected 
by the Analyzing Magnets to the beam-left side of the apparatus,
and those with the opposite sign to beam-right.
The highly redundant tracking system facilitated very high
track-reconstruction efficiencies.
The trigger required the coincidence of at least one hit counter
in each of the Same-Sign (SS) and Opposite-Sign (OS) hodoscopes situated 
on either side of the secondary beam (the LR subtrigger),
along with an energy deposit of at least ${\approx}40$\,GeV
in the hadronic calorimeter.
This energy threshold was well below the 60\,GeV of the 
lowest-energy protons from \Omm\ decays, all of which entered
the calorimeter, and above the energy where the 
calorimeter efficiency plateaued at ${\approx}99$\%.
Since there was a high probability that both the \Km\ and the \pim\
would hit the SS hodoscope and since the OS hodoscope
had two layers of counters, the efficiency of the LR subtrigger was
extremely high (${\approx}$99.5\%).
Events that satisfied the trigger were written to magnetic tape by
a high-rate data acquisition system \cite{daq}.

%\section{Event Selection}
The analysis reported here is from data taken with the 
negative-polarity secondary beam.
The 29 billion recorded events were initially reconstructed and separated
according to event type using loose event-selection cuts.
This left a total of 56 million candidate events.
The raw event information was preserved at this
(as well as every subsequent) stage.
Final event-selection criteria were applied after careful study
and were tuned to maximize the signal-to-background ratio.
The most important requirements were that:
(1) the $\chi^2/{\rm df}$ for a geometric fit to the
    decay topology be less than 2.5;
(2) the distance-of-closest-approach for
    the tracks forming the \gL\ and \Omm\ decay vertices be less
    than 4\,mm;
(3) the $x$ and $y$ separations from the target center of the 
    extrapolated \Omm\ trajectory satisfy the inequality
    $(x/2.0\,{\rm mm})^2 + (y/2.2\,{\rm mm})^2 \leq 1.0$;
(4) both the \Omm\ and the \gL\ decay vertices
    lie at least 0.28\,m (0.32\,m) downstream (upstream) of the
    entrance (exit) of the Vacuum Decay Region,
    and that the \Omm\ vertex precede that of the \gL;
(5) the \gp\pim\pim\ (\pip\pim\pim) invariant mass be 
    greater than 1.355\,GeV/$c^2$ (0.520\,GeV/$c^2$), in order to eliminate
    \Xim\ra\gL\pim\ra\gp\pim\pim\ (\Km\ra\pip\pim\pim) decays;
(6) the \gp\pim\ and \gp\pim\Km\ invariant masses be respectively
    within ${\pm}4.0$\,MeV/$c^2$ ($4.3\,\sigma$) and ${\pm}8.0$\,MeV/$c^2$
    ($5.0\,\sigma$) of the true \gL\ and \Omm\ masses; and
(7) no particle have momentum less than 12\,GeV/$c$.
After all these cuts the number of events remaining was 4.735 million.
Monte Carlo simulation indicated that 55.3\% of 
\Omm\ra\gL\Km\ra\gp\pim\Km\ decays for which the \Omm\ exited
the collimator passed these cuts.
The cuts that eliminated the greatest numbers of signal events 
were the \gp\pim\pim\ invariant mass and the \Omm\ vertex requirements.

\begin{figure}[htbp]
\centerline{\includegraphics[width=4.3in]{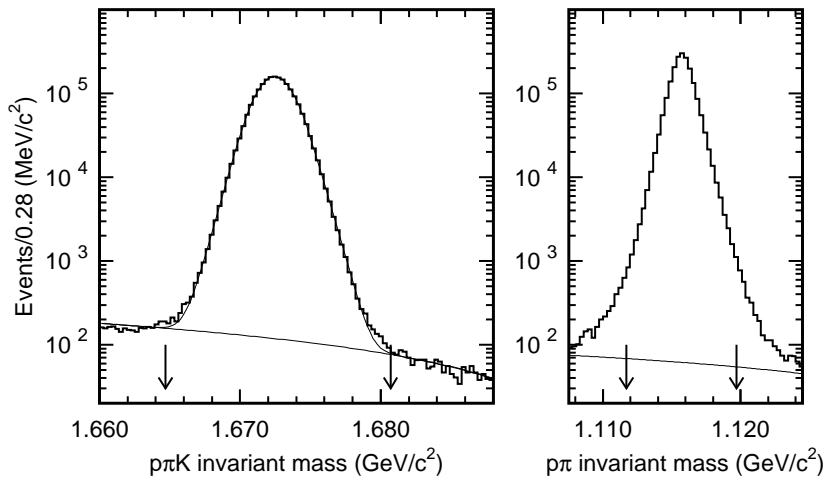}}
\caption{The \gp\pim\Km\ (left) and \gp\pim\ (right) invariant-mass
         distributions, after all cuts except the respective mass cuts.  
         Arrows delimit the accepted mass regions.
\label{fig:mass_omega}}
\end{figure}

Figure~\ref{fig:mass_omega} shows the \gp\pim\Km\ and \gp\pim\ 
invariant-mass distributions after event selection cuts. 
The background-to-signal ratio, determined using a double-Gaussian plus
second-degree polynomial fit to the invariant-mass distribution,
is $(0.33{\pm}0.03)$\% in the region within ${\pm}5.0\,\sigma$ of 
the \Omm\ mass.
The background under the \gp\pim\ mass peak is less than half this.
Dominant backgrounds were misreconstructed \Xim\ra\gL\pim\ra\gp\pim\pim\ 
decays and \Omm\ra\Xiz\pim \ra \gL\piz\pim \ra  \\
\gp \pim \pim \gam\gam\ decays.

\begin{figure}[htbp]
\centerline{\includegraphics[height=2.5in]{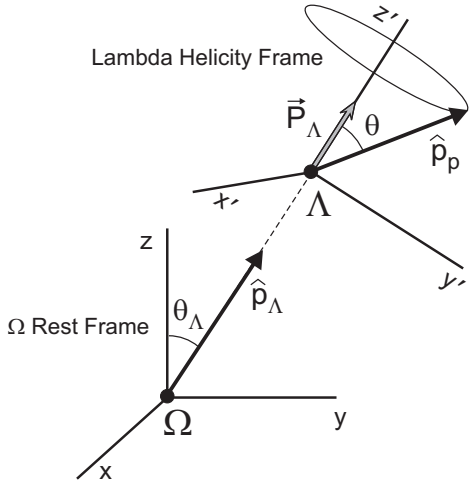}}
\caption{The Lambda Helicity Frame.
\label{fig:lhf}}
\end{figure}

%\section{Results}
The \Omm\ alpha parameter was measured through the asymmetry in the
\gL\ra\gp\pim\ decay distribution.
In the decay of an unpolarized \Omm\ to a \gL\ and a \Km, the 
\gL\ is produced in a helicity state with its helicity given 
by \alOm\ \cite{Kim}.
Hence the decay distribution of the proton in that \gL\ rest frame in
which the \gL\ direction in the \Omm\ rest frame defines the 
polar axis --- the Lambda Helicity Frame shown in 
Fig.~\ref{fig:lhf} --- is given by
\begin{equation}
   \frac{dN}{d\cos\theta} =
      \frac{N_0}{2}(1 + \alOm\alL\cos\theta),
\label{eq:2}
\end{equation}
where $\theta$ is the polar angle of the proton and
\alL\ is the alpha decay parameter in \gL\ra\gp\pim.
Since the \gL\ decay direction in the \Omm\ rest frame changes
from event to event, so too does the polar axis of the Lambda
Helicity Frame along which the \gL\ polarization must lie:
knowledge of the direction of the putative \gL\ polarization
is of enormous importance as it greatly minimizes biases.
The analysis ``locks in'' to the changing direction of the
\gL\ polarization.
Biases, on the other hand, such as uncorrected detector inefficiencies,
are fixed in the laboratory frame.
Hence the Lambda Helicity Frame analysis acts much like a lock-in amplifier, 
except that it locks into a known direction rather than a known frequency.

The proton $\cos\theta$ acceptance was measured and corrected for 
using a hybrid Monte Carlo (HMC) technique that has been used
in many similar such measurements \cite{HMC}.   
Monte Carlo events were generated by taking all parameters from real 
events except the proton and pion direction in the rest frame of the \gL.  
An isotropic \gL\ra\gp\pim\ decay was generated, and 
the proton and pion were boosted back into the laboratory frame using
the real \gL\ momentum.  Their trajectories were then traced through 
the apparatus, with the detector responses simulated where appropriate
(using measured efficiencies),
and all MWPC wire hits not associated with the real proton and pion
tracks kept.
The simulation included multiple scattering and slope-dependent
multiple-wire hit probabilities which were tuned to match data.
The HMC simulated the data extremely well, as is evident by
the small $\chi^2/{\rm df}$ in the matching of the real and
HMC proton distributions in the Lambda Helicity Frame (see discussion
below).  Real and HMC distributions of proton and pion tracks at various
places along the spectrometer also showed excellent agreement.
The HMC proton and pion tracks, in conjunction with the real kaon,
were required to satisfy the trigger requirements,
and were reconstructed by the standard 
track-finding program, with the same cuts applied to all parameters
formed from them that were applied to the real events.
Ten accepted HMC events were used for each real event.
If over 300 generated HMC events were required to get the ten, 
then both the real and associated HMC events were discarded;
this eliminated regions of low acceptance and reduced
the computer time needed for the analysis.  It eliminated
4.9\% of the events.
Increasing the upper limit beyond 300 events had no effect on the result.

Since the HMC events were generated with a uniform proton $\cos\theta$
distribution, each accepted HMC event was then weighted by 
\begin{equation}
      W = \frac{1 + S\cos\theta_f}{1 + S\cos\theta_r}, 
\label{eq:3}
\end{equation}
where $S$ is the slope (to be determined) of the proton $\cos\theta$ 
distribution and $\theta_f$ and $\theta_r$ are, respectively, the HMC 
(``fake'') and real proton polar angles in 
the Lambda Helicity Frame.
Note that in the absence of a background correction $S = \alOm\alL$.
The numerator in Eq.~(\ref{eq:3}) in effect polarizes the HMC sample,
while the denominator removes the polarization bias accrued from
using parameters from real polarized \gL\ decays. 
To facilitate handling the unknown slope $S$, the weights, 
binned in $\cos\theta_f$, were approximated by the 
polynomial series expansion
\begin{equation}
      W \approx (1 + S\cos\theta_f)[1 - S\cos\theta_r +
                       (S\cos\theta_r)^2 - \cdots + (S\cos\theta_r)^{10}].
\label{eq:4}
\end{equation}
The polynomial coefficients, which depend only
on $\cos\theta_f$ and $\cos\theta_r$, were summed, and then $S$ was 
extracted by minimizing the $\chi^2$ difference 
between the real and weighted HMC proton $\cos\theta$ distributions.
The error was determined by finding
the variation in $S$ needed to increase $\chi^2$ by one.
It includes the uncertainty in the acceptance as determined
by the HMC events.

The analysis procedure was extensively checked by Monte Carlo simulation.
Monte Carlo \Omm\ra\gL\Km\ra\gp\pim\Km\ events were simulated using the 
real hodoscope, MWPC, and calorimeter efficiencies,
and required to pass the same cuts as the real data. These were analyzed
by the HMC analysis code.  
The input and extracted values of \alOm\alL\ were found to be consistent
over a wide range of \alOm\ input values, with an
average difference of $(0.017{\pm}0.042){\times}10^{-2}$. 
As a cross-check, 78\,000 \Xim\ra\gL\pim\ra\gp\pim\pim\ decays 
available from the same dataset were analyzed using exactly 
the same analysis program, with selection criteria tuned for the \Xim\ decay.
The fit to the proton $\cos\theta$ distribution was good,
with $\chi^2/{\rm df} = 14/19$.
The correct sign of \alXi\alL\ was found, which is opposite the sign
of our value of \alOm\alL, and the magnitudes
of the measured and PDG values of \alXi\alL\ were consistent
within the statistical errors.
\begin{figure}[hbtp]
\centerline{\includegraphics[width=3.1in]{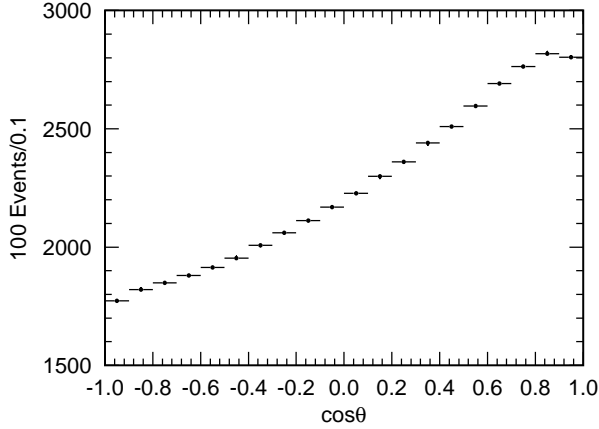}}
\caption{The real (lines) and weighted HMC (points) proton 
         $\cos\theta$ distributions.   The total number of 
         HMC events has been scaled down by a factor of 10.
\label{fig:cost_raw_rwmc}}
\end{figure}

\begin{figure}[htbp]
\centerline{\includegraphics[width=3.4in]{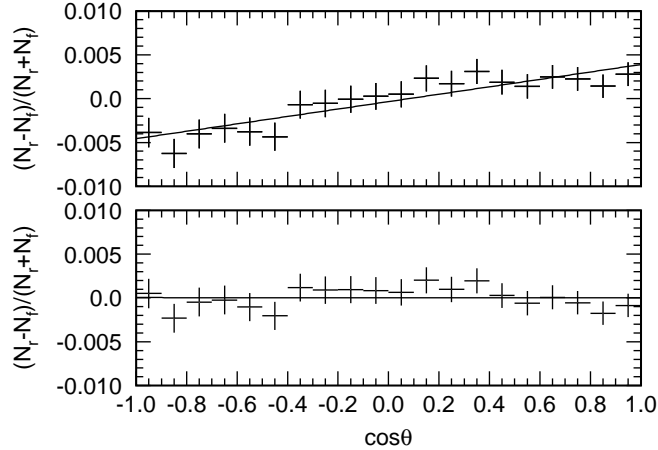}}
\caption{The relative differences between the real (N$_{\rm r}$) 
         and HMC (N$_{\rm f}$) proton $\cos\theta$ distributions
         for unweighted (top) and weighted (bottom) HMC events.
         The total number of HMC events has been scaled down by a factor of 10.
\label{fig:cost_rf_wgt_unwgt}}
\end{figure}

A total of 4\,504\,896 real events were analyzed by the method described
above.  The real and weighted HMC proton $\cos\theta$ distributions are
shown in Fig.~\ref{fig:cost_raw_rwmc}, and 
the differences between the real and HMC proton $\cos\theta$ distributions, 
weighted and unweighted, are shown in Fig.~\ref{fig:cost_rf_wgt_unwgt}.
The nonisotropic nature of the real proton $\cos\theta$ distribution,
compared to the isotropically generated HMC distribution, is clear
from the top plot of Fig.~\ref{fig:cost_rf_wgt_unwgt}.
It is unambiguous evidence of a nonzero $\alpha$ decay parameter.
The bottom plot shows the same comparison, except that the HMC events
have been weighted by the best-fit value of $S$.
The extracted slope of the proton $\cos\theta$ distribution is
$S = (1.16{\pm}0.12){\times}10^{-2}$ with $\chi^2/{\rm df} = 23/19$,
where the error is statistical.

To extract \alOm\alL\ from the proton $\cos\theta$ slope,
the small background contribution to $S$ was subtracted.
To estimate the proton $\cos\theta$ slope from the background events the same
analysis procedure was performed on five sideband regions, 
three below and two above the \Omm\ mass region.
The average sideband proton $\cos\theta$ slope was found to be
$S_{\rm sb} = (7.2{\pm}3.0){\times}10^{-2}$, with
average $\chi^2/{\rm df} = 19/19$.
No mass dependence of $S_{\rm sb}$ was apparent.
The contribution to $S$ of the background under the mass peak was
corrected for by subtracting the appropriate fraction of 
$S_{\rm sb}$ from $S$, giving
$\alOm\alL = [1.14{\pm}0.12({\rm stat})]{\times}10^{-2}$.
Note that this correction is only a 1.7\% effect.
\begin{figure}[htbp]
\centerline{\includegraphics[width=3.4in]{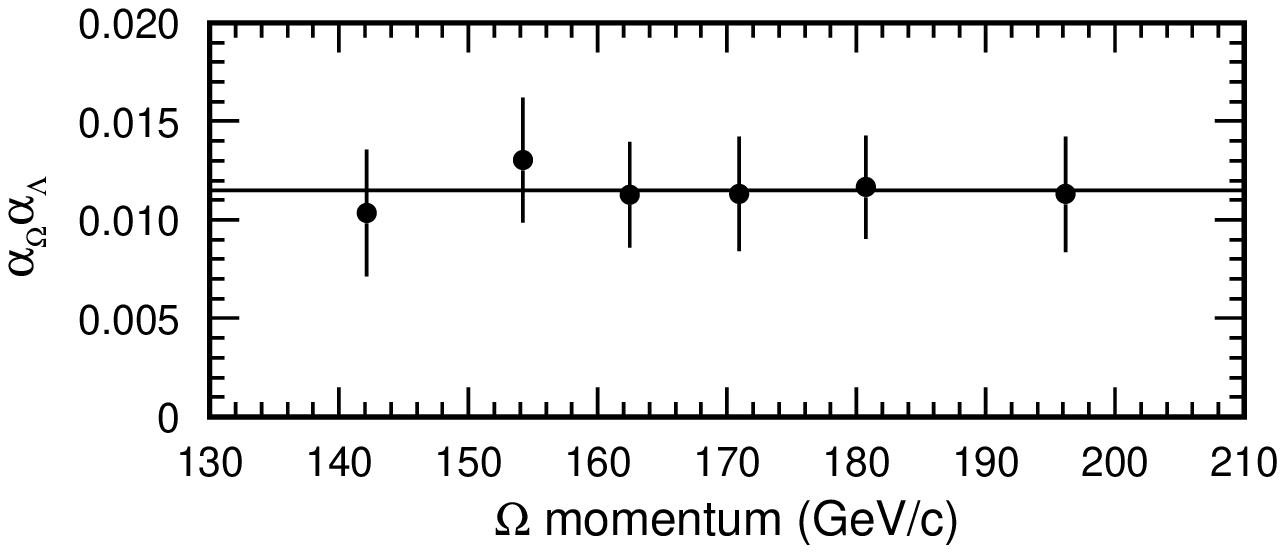}}
\caption{The value of \protect\alOm\protect\alL\ vs.\ the \protect\Omm\ momentum.
\label{fig:alpha_vs_p}}
\end{figure}

\begin{figure}[htbp]
\centerline{\includegraphics[width=3.4in]{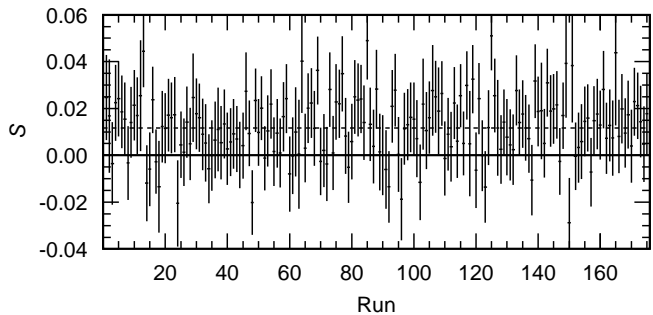}}
\caption{Run-by-run values of the proton slope $S$.  The dashed line
represents the best-fit value.
\label{fig:s_vs_run}}
\end{figure}

The stability of the result was studied as a function of several parameters.
The value of \alOm\alL\ was independent of the \Omm\ momentum,
as shown in Fig.~\ref{fig:alpha_vs_p}, 
and there was no dependence on the $z$ location of the \Omm\ decay vertex.  
The non-background-subtracted slope $S$, measured on a run-by-run basis
for all 175 runs in the dataset, shows no evidence of a 
temporal dependence (see Fig.~\ref{fig:s_vs_run}).

Systematic errors were small because of the high efficiencies
of the spectrometer elements and because of the power of the 
Lambda Helicity Frame analysis.  
The dominant systematic errors are listed
in Table~\ref{tab:systematics}.
The effects of uncertainties in detector inefficiencies --- MWPCs, trigger
hodoscopes, and hadronic calorimeter --- on \alOm\alL\ were
found to be negligible.
No statistically significant difference in $S$ was found 
between using perfect and measured detector efficiencies when 
simulating the HMC proton and pion.
The combined effect of the uncertainties in the fields of the Analyzing Magnets,
5.5\,G, was also negligible.
A small fraction of the daughter \pim's and \Km's decayed before
exiting the apparatus.  (Approximately 0.7\% of the \pim's decayed before
the last MWPC.)  The effect of such decays on \alOm\alL\
was studied using Monte Carlo events and data and found to be negligible.
The error in the background subtraction was estimated by assuming that
the error in the average sideband slope $S_{\rm sb}$ was equal to the average
sideband slope, ${\delta}S_{\rm sb} = 7.2{\times}10^{-2}$, 
and using a 25\% error in the background-to-signal ratio,
both very conservative assumptions.  It too is negligible.
\begin{table}[htb]
\renewcommand{\arraystretch}{1.1}
\begin{center}
\caption{Systematic errors. \label{tab:systematics}}
\vspace*{0.1in}
\begin{tabular}{lr@{.}l}
  \hline\hline
  \multicolumn{3}{l}{\hspace*{1.1in}Source\hspace*{1.3in}Error ($10^{-2}$)} \\
  \hline
  Event selection cut variations                              &  0 & 088  \\
  Validation of analysis code                                 &  0 & 042  \\
  Background subtraction uncertainty                          &  0 & 024  \\
  Detector inefficiency uncertainties                         &  0 & 010  \\
  Analyzing Magnets field uncertainties \hspace*{0.6in}       &  0 & 006  \\
\hline\hline
\end{tabular}
\end{center}
\end{table}

The largest systematic uncertainty was the sensitivity
of the measurement to the values of the cuts used to define the 
data sample.  The most important were the 
cuts on the \gp\pim\ and \gp\pim\Km\ invariant masses and,
less importantly, the 12\,GeV/$c$ minimum momentum cut.
The effect of changes in these cut values
was $0.088{\times}10^{-2}$.
The total systematic error, including the upper limit in the
uncertainty of the MC validation of the analysis program 
($0.042{\times}10^{-2}$), is $0.10{\times}10^{-2}$.
This is a factor of five reduction in the systematic error of 
$0.52{\times}10^{-2}$ reported in the analysis of the 1997 
data \cite{hypercp97};
most of the improvement comes from incorporating the measured detector and 
track-finding inefficiencies into the HMC simulation in this analysis.

To summarize, we find from a sample of 4\,504\,896 
\Omm\ra\gL\Km\ra\gp\pim\Km\ decays the value
$\alOm\alL = [1.14{\pm}0.12({\rm stat}){\pm}0.10({\rm syst})]{\times}10^{-2}$.
Using $\alL = 0.642{\pm}0.013$ \cite{pdg}, \alOm\ is 
found to be 
$[1.78{\pm}0.19({\rm stat}){\pm}0.16({\rm syst})]{\times}10^{-2}$,
where the contribution of the uncertainty in the value of \alL\ to the
systematic error is negligible.
Our measurement represents a factor of nine improvement in precision
over the current PDG value.  It is $1.9\,\sigma$ from the PDG average 
of $(-2.6{\pm}2.3){\times}10^{-2}$, and opposite in sign.
This measurement is consistent with the recent result we reported
\cite{hypercp97} from an independent analysis of data taken in the 
1997 fixed-target running period, but with a factor of four
smaller error. 
With a magnitude that is 7.2\,$\sigma$ from zero, 
it represents unambiguous evidence
of parity violation in the \Omm\ra\gL\Km\ decay.
As predicted, \alOm\ is small; indeed it is
the smallest of all the $\alpha$ parameters that have been measured in the
two-body weak decays of hyperons.

\section*{Acknowledgments}
The authors are indebted to the staffs of Fermilab and the
participating institutions for their vital contributions.
This work was supported by the U.S. Department of Energy
and the National Science Council of Taiwan, R.O.C.
E.C.D. and K.S.N. were partially supported by the
Institute for Nuclear and Particle Physics at the University 
of Virginia.
K.B.L. was partially supported by the Miller Institute for
Basic Research in Science.

\end{document}